\documentclass[10pt]{iopart}
\usepackage{graphicx}
\usepackage{setstack}
\usepackage{amssymb}

\begin{document}
\title[Ferrimagnetism in Mn$_2$TiZ Heusler compounds]{\textit{Ab initio} prediction of ferrimagnetism, exchange interactions and Curie temperatures in Mn$_2$TiZ Heusler compounds}
\author{M Meinert, J M Schmalhorst, and G Reiss}
\address{Department of Physics, Bielefeld University,
33501 Bielefeld, Germany}
\ead{meinert@physik.uni-bielefeld.de}

\date{\today}

\begin{abstract}
The Heusler compounds Mn$_2$Ti\textit{Z} (\textit{Z} = Al, Ga, In, Si, Ge, Sn, P, As, Sb) are of large interest due to their potential ferrimagnetic properties and high spin polarization. Here, we present calculations of the structural and magnetic properties of these materials. Their magnetic moment follows the Slater-Pauling rule $m = N_V - 24$. None of them is actually a perfect half-metallic ferrimagnet, but some exhibit more than 90\% spin polarization and Curie temperatures well above room temperature. The exchange interactions are complex, direct and indirect exchange contributions are identified. The Curie temperature scales with the total magnetic moment, and it has a positive pressure dependence. The role of the \textit{Z} element is investigated: it influences the properties of the compounds mainly via its valence electron number and its atomic radius, which determines the lattice parameter. Based on these results, Mn$_2$TiSi, Mn$_2$TiGe, and Mn$_2$TiSn are proposed as candidates for spintronic applications.
\end{abstract}

\maketitle

\section{Introduction}
A very interesting class of Heusler compounds that has received considerable theoretical, but only few experimental attention to date, are the half-metallic ferrimagnets Mn$_2$\textit{YZ}, where \textit{Y} = V, Cr, Mn, Fe, Co, Ni, Cu and \textit{Z} is a group III, IV, or V element \cite{Oezdogan06, Fujii08, Luo08_1, Wurmehl06, Luo08_2, Liu08, Xing08, Luo09, Wei10}. Half-metallic compounds are characterized by a gap for either the spin-down or the spin-up density of states (DOS) at the Fermi energy, so that an electric current has purely up or down electrons. This property makes them highly interesting for applications in spintronics. A half-metallic ferrimagnet has advantages over the well-known half-metallic ferromagnets: due to the internal spin compensation it has rather low magnetic moment, while the Curie temperature remains fairly high. A low magnetic moment gives rise to low stray fields, which is desired for spintronics, as is a high Curie temperature and thus a good thermal stability of the compound \cite{Pickett01}. The most prominent compound out of this class is Mn$_2$VAl, which has been studied thoroughly by experiment and theory \cite{Itoh83,Yoshida81,Jiang01,Ishida84,Weht99}. Together with numerous other compounds in the Mn$_2$V\textit{Z} series it has been predicted to be a half-metallic ferrimagnet \cite{Oezdogan06,Sasioglu05_1}. Its low magnetic moment of about $2\,\mu_\mathrm{B}$ per formula unit (f.u.) and the high Curie temperature of 760\,K make it a promising compound for spintronics \cite{Jiang01}. Several other materials classes have been proposed to be half-metallic ferrimagnets, e.g., Cr$_{0.75}$Mn$_{0.25}$Se and Cr$_{0.75}$Mn$_{0.25}$Te in the zinc blende structure \cite{Nakamura05}, or Cr antisites in CrAs, CrSb, CrSe, and CrTe, having the zinc blende structure \cite{Galanakis06}.

Ideally, an electrode material for spintronics would be a half-metal with zero net moment. This can not be achieved with antiferromagnets because of the spin-rotational symmetry (resulting in zero polarization), but well chosen half-metallic ferrimagnets can be tuned to zero moment. This property is also known as half-metallic antiferromagnetism, and has been first predicted for Mn and In doped FeVSb \cite{vanLeuken95}. Among others, La$_2$VMnO$_6$ and related double perovskites \cite{Pickett98} and certain diluted magnetic semiconductors have been later predicted to be half-metallic antiferromagnets as well \cite{Akai06}. Finally, the ferrimagnetic Heusler compounds Mn$_2$VAl and Mn$_2$VSi have been proposed as a starting point for doping with Co to achieve the full compensation \cite{Galanakis07}. However, it should be noted that the half-metallic antiferromagnetism is limited to zero temperature and a small macroscopic net moment is expected at elevated temperature---in particular near the Curie temperature---because of the inequivalent magnetic sublattices \cite{Sasioglu09}.

Following the Slater-Pauling rule connecting the magnetic moment $m$ and the number of valence electrons $N_V$ via $m = N_V - 24$ in the half-metallic Heusler compounds \cite{Galanakis02}, it is expected to find another series of ferrimagnetic half-metals in the Mn$_2$Ti\textit{Z} system with $-3$ to $-1 \,\mu_\mathrm{B}$\,/\,f.u. The negative moment indicates that the half-metallic gap would appear for the majority states. These compounds could---if they are half-metals---provide another series of potential electrodes for spin-dependent applications and could also become a starting point for half-metallic antiferromagnetism.

In this paper, we discuss \textit{ab initio} calculations of the properties of the (hypothetical) Mn$_2$Ti\textit{Z} compounds, crystallized in the L2$_1$ structure. No experimental data are available for this system, and only Mn$_2$TiAl has been studied theoretically before \cite{Luo08_3}. However, it is expected that parts of this series will exist in the L2$_1$ structure, seeing that Mn$_2$VAl and Mn$_2$VGa, as well as parts of the Co$_2$Ti\textit{Z} series have been prepared \cite{Kumar08, Barth10, Graf10}.

\section{Calculational approach}
The calculations presented in this study were performed within two different density functional theory-based band structure codes: the full-potential linearized augmented plane waves (FLAPW) package Elk \cite{elk} and the full-potential Korringa-Kohn-Rostoker \textit{Munich} SPRKKR \cite{sprkkr} package. Although both methods are in principle equivalent for crystalline systems, there are subtle differences associated with their numerical implementations, and thus it is worth to compare both methods on the rather complex intermetallic system Mn$_2$TiZ.

Elk was used to determine the theoretical lattice parameters and the total energy differences between ferrimagnetic and nonmagnetic states. These calculations were carried out on a $12 \times 12 \times 12$ $\bi{k}$ point mesh (72 points in the irreducible wedge of the Brillouin zone). The muffin-tin radii of all atoms were set to 2.0 a.u. to avoid overlaps at small lattice parameters. The equilibrium lattice parameters $a$ were determined using a third-degree polynomial fit to the total energies. To obtain accurate magnetic moments and densities of states, the calculations were performed at the equilibrium lattice parameter using a $16 \times 16 \times 16$ $\bi{k}$-mesh (145 points in the irreducible wedge) and nearly touching muffin-tin spheres.

The SPRKKR calculations were performed on the theoretical equilibrium lattice parameters determined with Elk. The calculations were carried out in the full-potential mode with an angular momentum cutoff of $l_{max} = 3$ on a $22 \times 22 \times 22$ $\bi{k}$ point mesh (289 points in the irreducible wedge of the Brillouin zone). Both the full potential as well as the increased angular momentum cutoff are necessary to ensure accurate results. The DOS were calculated on a denser mesh of 1145 $\bi{k}$ points with 0.5\,mRy added as the imaginary part to the energy.

The exchange-correlation potential was modeled within the generalized gradient approximation of Perdew, Burke, and Ernzerhof in both schemes \cite{PBE}. The calculations were converged to about 0.1 meV. All calculations were carried out in the scalar-relativistic representation of the valence states, thus neglecting the spin-orbit coupling.

\begin{figure*}[t]
\begin{center}
\includegraphics[scale=1]{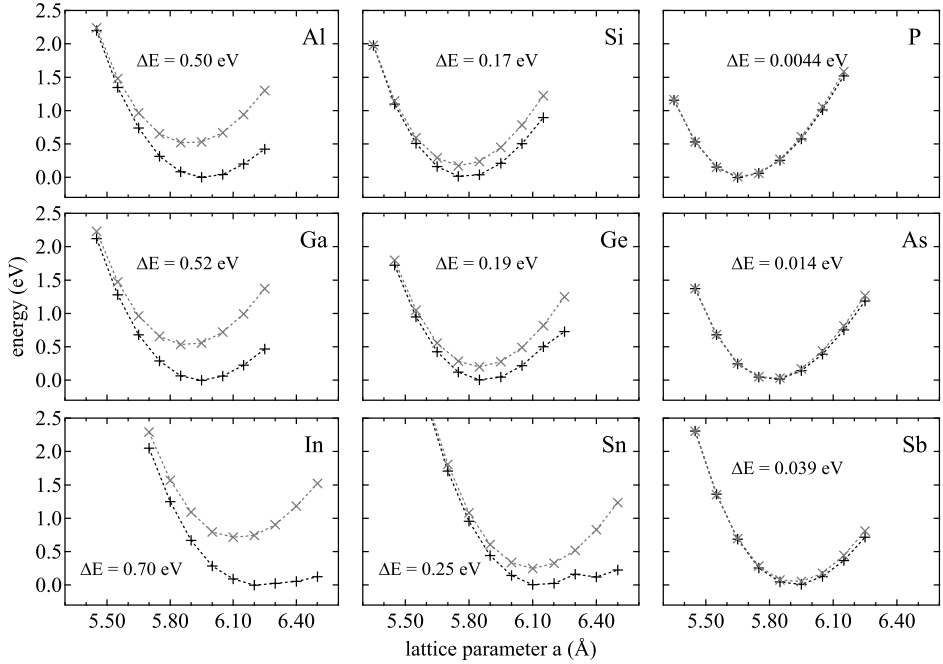}
\end{center}
\caption{\label{Fig1}Total energies of the investigated compounds in dependence of their lattice parameters. The results for the ferrimagnetic and the non-magnetic states are represented with $+$ and $\times$, respectively.}
\end{figure*}

SPRKKR allows to calculate the Heisenberg exchange coupling parameters $J_{ij}$ within a real-space approach using an expression proposed by Liechtenstein \textit{et al.} \cite{Liechtenstein87}. Using the $J_{ij}$ the Curie temperatures were calculated within the mean field approximation (MFA). For a single-lattice system the Curie temperature is given within the MFA by
\begin{equation}\label{eq:single}
\frac{3}{2} k_\mathrm{B} T_\mathrm{C}^{\mathrm{MFA}} = J_0 = \sum\limits_j J_{0j}.
\end{equation}
In a multi-sublattice system---as, e.g., the Heusler compounds with four sublattices---one has to solve the coupled equations
\begin{eqnarray}\label{eq:multi}
\frac{3}{2} k_\mathrm{B} T_\mathrm{C}^{\mathrm{MFA}} \left< e^\mu \right> &=& \sum\limits_\nu J_0^{\mu \nu} \left< e^\nu \right>\\
J_0^{\mu \nu} &=& \sum\limits_{\bi{R}\neq 0} J^{\mu \nu}_{0 \bi{R}}, \nonumber
\end{eqnarray}
where $\left< e^\nu \right>$ is the average $z$ component of the unit vector $e_\bi{R}^\nu$ pointing in the direction of the magnetic moment at site ($\nu$, $\bi{R}$). The coupled equations can be rewritten as an eigenvalue problem:
\begin{eqnarray}\label{eq:eigen}
(\bi{\Theta} - T\, \bi{I})\,\bi{E} &=& 0\\
\frac{3}{2} k_\mathrm{B} \Theta_{\mu \nu} &=& J_0^{\mu \nu}
\nonumber
\end{eqnarray}
with a unit matrix $\bi{I}$ and the vector $E^\nu = \left< e^\nu \right>$. The largest eigenvalue of the $\bi{\Theta}$ matrix gives the Curie temperature \cite{Sasioglu05_1, Anderson63}. In order to separate the two Mn lattices, the calculations were run in F$\bar{4}3$m space group, in which the Mn atoms are not equivalent by symmetry. The $\bi{R}$-summation in Eq. (\ref{eq:multi}) was taken to a radius of $R_\mathrm{max} = 3.0\,a$, which has been shown to be sufficient for half-metallic Heusler compounds \cite{Rusz06, Thoene09}.

\section{Results}

\subsection{Energy minimization and lattice parameters}
Three types of magnetic configurations were tested: ferro-, ferri-, and nonmagnetic. It was found for all compounds that the ferromagnetic configurations were unstable and converged into the ferrimagnetic state. Fig. \ref{Fig1} displays the total energies of the ferrimagnetic and the nonmagnetic configurations in dependence on the lattice parameters $a$. We find that the ferrimagnetic state has always lower energy than the non-magnetic state; the difference in total energy reduces with increasing number of valence electrons, but it increases within the groups with the atomic number. The lattice parameters follow roughly a linear dependence on the atomic radius of the \textit{Z} element with the correlation coefficient of $r=0.92$ (Fig. \ref{Fig2} (a)). Some compounds show a strong asymmetry of the total energy curve in the ferrimagnetic configuration and even kinks in the curves for very large $a$. This is caused by a steep increase of the magnetic moments for increasing $a$ which causes a stronger binding. However, this effect is never strong enough to shift the equilibrium lattice parameter to such a high-$m$ state. The equilibrium lattice parameters are summarized in Table \ref{overview}. Typically we find the equilibrium lattice parameters of Heusler compounds obtained with Elk to be accurate within $\pm 0.5\,\%$ compared to experiment.

\fulltable{\label{overview} Results of the ground state properties calculations with Elk and SPRKKR. The total magnetic moments are given in $\mu_B$ per formula unit, the atomic magnetic moments are given in $\mu_B$ per atom. The SPRKKR results for Mn$_2$TiAs were obtained with $a = 5.95$\,\AA{} (see text). }
\br
		& \centre{5}{Elk}	&  \centre{4}{SPRKKR}	 \\
\ns\ns		& \crule{5}		&  \crule{4}			\\
Mn$_2$Ti\textit{Z}	& $a$ (\AA{})	 	&	$m$		& $m_\mathrm{Mn}$ 	& $m_\mathrm{Ti}$	& P (\%)	&	$m$		& $m_\mathrm{Mn}$ 	& $m_\mathrm{Ti}$	& P (\%)	\\ \mr
Al	& 5.96		&	2.98		& 1.83		& -0.57		& 21		& 2.98		& 1.76		& -0.49	& 82		\\
Ga	& 5.95		&	2.95		& 1.84		& -0.60		& 45		& 2.97		& 1.77		& -0.53	& 79		\\
In	& 6.23		&	3.17		& 2.17		& -0.86		& 7 		& 3.08		& 1.98		& -0.82	& 32		\\ \bs

Si	& 5.78		&	1.98		& 1.16		& -0.31		& 94		& 1.98		& 1.13		& -0.26	& 87		\\
Ge	& 5.87		&	1.97		& 1.20		& -0.37		& 94		& 1.97		& 1.16		& -0.33	& 89		\\
Sn	& 6.14		&	1.97		& 1.32		& -0.51		& 97		& 2.00		& 1.25		& -0.48	& 93		\\ \bs

P	& 5.68		&	0.30		& 0.18		& -0.05	& -3		& ---		& ---		& ---		& ---		\\
As	& 5.82		&	0.94		& 0.59		& -0.20 	& 84		& 0.97		& 0.61		& -0.22	& 58		\\
Sb	& 6.07		&	0.97		& 0.65		& -0.25 	& 88		& 0.98		& 0.62		& -0.24	& 79		\\
\br
\endfulltable

\subsection{Magnetic moments and densities of states}

The results discussed in this subsection are summarized in Table \ref{overview} and Fig. \ref{Fig3}.

\subsubsection{Mn$_2$TiAl, Mn$_2$TiGa, Mn$_2$TiIn}

From the rule $m = N_V - 24$ we expect to find a magnetic moment of $3\,\mu_\mathrm{B} / \mathrm{f.u.}$ for these compounds. The FLAPW calculations show small deviations from this rule, indicating that the compounds are not perfect half-metals. This is confirmed by the DOS, which show spin polarizations at the Fermi level below 50\,\%, and in particular only 7\,\% for Mn$_2$TiIn, where the magnetic moment is enhanced to $3.17\,\mu_\mathrm{B} / \mathrm{f.u.}$. This arises from the large lattice parameter and the fact that all three compounds do not form a gap in the DOS. The Fermi level for Mn$_2$TiAl and Mn$_2$TiGa is in a region with low DOS for both spin channels (see insets in Fig. \ref{Fig3}), but both of them have a very large empty spin-down DOS right above $E_\mathrm{F}$. Small variations of the lattice parameter would thus lead to strong variations of the spin polarization.

\begin{figure}[t]
\begin{center}
\includegraphics[scale=1]{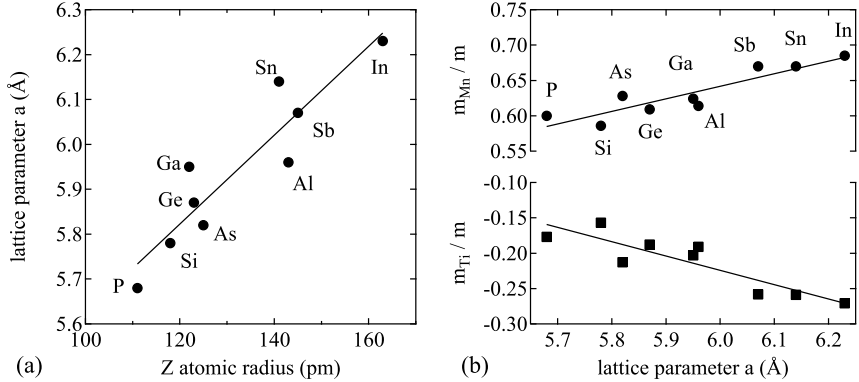}
\end{center}
\caption{\label{Fig2}(a): Dependence of the lattice parameter $a$ on the atomic radius of the \textit{Z} element. (b): Normalized magnetic moments of Mn and Ti in dependence of the lattice parameter.}
\end{figure}

The calculations performed with SPRKKR reproduce the magnetic moments obtained in Elk very well. Although the total moments are practically equal, a larger deviation is found for the atom-resolved moments. The Fermi energy is found at slightly different positions in the DOS, and the detailed structures observed in Elk around $E_\mathrm{F}$ are less pronounced, especially the dip in the spin-down states at $E_\mathrm{F}$. This leads to significantly higher spin polarization values in SPRKKR. However, the trend that Mn$_2$TiIn has the lowest polarization within this group is reproduced.

\begin{figure*}[t]
\begin{center}
\includegraphics[scale=1]{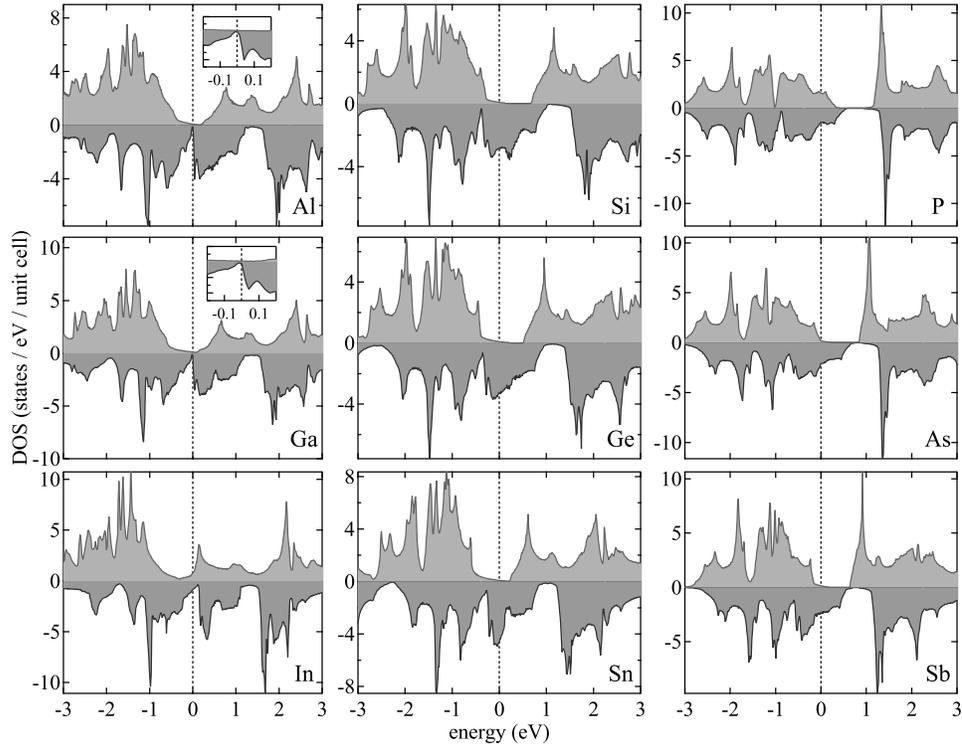}
\end{center}
\caption{\label{Fig3}Densities of states calculated with Elk. The spin-up DOS is pointing up, the spin-down DOS is pointing down. The insets for Al and Ga show the region around the Fermi energy.}

\end{figure*}

\subsubsection{Mn$_2$TiSi, Mn$_2$TiGe, Mn$_2$TiSn}

According to the ``rule of 24'' a total magnetic moment of $2\,\mu_\mathrm{B} / \mathrm{f.u.}$ is expected. Again, small deviations from this rule are observed; all moments are lower by about 1.5\,\%. In Elk, the three compounds are found to form a half-metallic gap in the spin-up states slightly above $E_\mathrm{F}$. The gap onset above $E_\mathrm{F}$ (width) is 0.16\,eV (0.49\,eV) for Si, 0.24\,eV (0.25\,eV) for Ge, and 0.19\,eV (0.01\,eV) for Sn.  Nevertheless, the spin polarization is above 90\,\% in these calculations. The structure of the DOS around $E_\mathrm{F}$ leads to a stable spin polarization and magnetic moment upon isotropic lattice compression or expansion. For this series, having the same valence electron counts and nearly half-metallic DOS, one can observe clearly a narrowing of the bands, i.e., the DOS are contracted towards $E_\mathrm{F}$, while the Fermi level itself moves upwards. This is directly associated with the gradually increasing lattice parameter in this series, which reduces the overlap of the 3d orbitals and thereby reduces the itinerancy of the system. An increased localization of the electrons provides also an explanation for the increasing atomic magnetic moments along this series. Similar behavior has been observed earlier for Co$_2$Mn\textit{Z}, with \textit{Z}\,=\,Si, Ge, Sn \cite{Picozzi02, Kurtulus05} and Ni$_2$MnSn \cite{Sasioglu05_2}. In the first case the Mn moment is increased and the Co moment is lowered along the series, keeping the total moment integer. Calculations on Co$_2$MnSi with increased lattice parameter reproduced this behavior. In the second case, the pressure dependence of the moments was studied. Under increasing pressure, i.e., with reduced lattice parameter, both the Ni and the Mn moment decrease, and thus the total moment decreases. However, Ni$_2$MnSn is not a half-metal, hence the total moment is not restricted to an integer value. Consequently, both observations on quite different ferromagnetic Heusler compounds are in accord with our case of (nearly) half-metallic ferrimagnetic Heusler compounds.

We note, that the magnetic moments and DOS from SPRKKR are in very good agreement with the ones obtained from Elk. However, the Fermi level is found at a lower position, giving rise to the slightly reduced polarization values.

\subsubsection{Mn$_2$TiP, Mn$_2$TiAs, Mn$_2$TiSb}

In these cases a total magnetic moment of only $1\,\mu_\mathrm{B} / \mathrm{f.u.}$ is expected. Because of the very small lattice parameter of Mn$_2$TiP, its spin-splitting is small with only $0.3\,\mu_\mathrm{B} / \mathrm{f.u.}$ in the Elk calculation. The situation of Mn$_2$TiAs and Mn$_2$TiSb is similar to that of Mn$_2$TiSi and Mn$_2$TiGe. A spin-up gap is formed  above the Fermi level with onset (width) of 0.29\,eV (0.53\,eV) for As and 0.19\,eV (0.44\,eV) for Sb. Though not being half-metallic, both compounds have spin polarizations of more than 80\,\%.

Finally, the magnetic moments of Mn$_2$TiSb in SPRKKR agree very well with those obtained with Elk. But again, the Fermi level is lower and the spin polarization is reduced. For Mn$_2$TiP and Mn$_2$TiAs the situation is quite different. They can not be converged into ferrimagnetic states at the equilibrium lattice parameters determined by Elk; instead, they are found to be nonmagnetic. This is caused by the tiny energy difference between the ferrimagnetic and the nonmagnetic configuration, which leads to a numerical instability of the ferrimagnetic state. By increasing the lattice parameter of Mn$_2$TiAs by about 2\,\% to 5.95\,\AA{}, the separation is increased artificially to about 30\,meV/f.u. and the calculation converges into the ferrimagnetic state. Because of this, the properties obtained with SPRKKR for this compound have to be taken with care: in all other cases the individual atomic moments are slightly lower in SPRKKR than those from Elk; here instead, larger moments are found. However, the same procedure can not be applied to Mn$_2$TiP, within a reasonable range of lattice parameters.

\subsubsection{General remarks}

It is worth to note that the magnetic moments of the $Z$ component are always below 0.06\,$\mu_\mathrm{B}$ and that they are always parallel to the Ti moment. In detail, the values are Al 0.044\,$\mu_\mathrm{B}$, Ga 0.052\,$\mu_\mathrm{B}$, In 0.058\,$\mu_\mathrm{B}$, Si 0.034\,$\mu_\mathrm{B}$, Ge 0.035\,$\mu_\mathrm{B}$, Sn 0.034\,$\mu_\mathrm{B}$, P 0.0062\,$\mu_\mathrm{B}$, As 0.018\,$\mu_\mathrm{B}$, and Sb 0.017\,$\mu_\mathrm{B}$.

Another property worth noting is the fact that the ratios $m_\mathrm{Mn}/m$ and $m_\mathrm{Ti}/m$ follow a linear dependence (with correlation coefficients of $r \approx 0.9$ in both cases for the Elk data) on the lattice parameter (and hence the interatomic distances) independently on the $Z$ type, see Fig. \ref{Fig2} (b). As mentioned above, with increasing lattice parameter the itinerant character of the system is reduced and localizes the moments gradually on the atoms. Therefore, the influence of the \textit{Z} component in Mn$_2$Ti\textit{Z} is twofold. First, it determines the lattice parameter of the compound and following from that, the degree of electron localization. And second, the total magnetic moment is determined via the number of electrons supplied, if the lattice parameter does not exceed a certain range (which is not the case for P and In).

\subsection{Exchange interactions and Curie temperatures}

\begin{figure}[t]
\begin{center}
\includegraphics[scale=1]{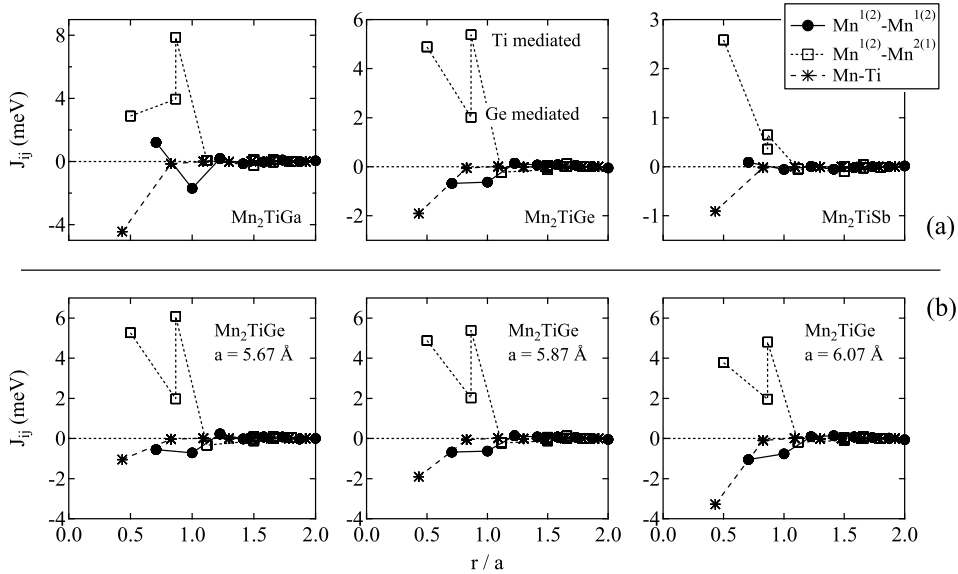}
\end{center}
\caption{\label{Fig4}Heisenberg exchange parameters $J_{ij}$ in dependence on the normalized distance $r/a$. (a): $J_{ij}$ for Mn$_2$TiGa, Mn$_2$TiGe, Mn$_2$TiSb for their respective equilibrium lattice parameters. (b): $J_{ij}$ for Mn$_2$TiGe with different lattice parameters. Note the different scales of the vertical axes in the top row.}
\end{figure}

The exchange interactions are investigated here for Mn$_2$TiGa, Mn$_2$TiGe, and Mn$_2$TiSb, which are representative compounds for their respective \textit{Z} group. Fig. \ref{Fig4} (a) displays the $J_{ij}$ calculated for the intra-sublattice interaction Mn$^{1(2)}$-Mn$^{1(2)}$ and the inter-sublattice interactions Mn$^{1(2)}$-Mn$^{2(1)}$ and Mn-Ti of the three compounds. All other interactions are very small and can be neglected for the following discussion.

In all three cases it is clear that the Mn$^{1(2)}$-Mn$^{2(1)}$ inter-sublattice interaction provides the largest contribution to the exchange. Further, the nearest neighbor interaction of Mn-Ti is always negative, hence all compounds are ferrimagnets. All interactions are mostly confined within a radius of $1.5\,a$. Apart from these similarities, there are many interesting differences.

First, we discuss the details of the dominating inter-sublattice interaction Mn$^{1(2)}$-Mn$^{2(1)}$. The first and second nearest neighbors provide a large, positive exchange. The second nearest neighbors have two different values of $J_{ij}$. This is a feature that is not observed in frozen-magnon calculations (see, e.g. \cite{Sasioglu05_1}), because the Fourier transform that is necessary to obtain the exchange parameters involves a spherical averaging. Instead, with the real-space approach used here we observe a difference for Mn atoms with a Ti atom or a \textit{Z} atom in between. We found larger values on the Mn atoms mediated via Ti and lower values on the \textit{Z} mediated ones. The nearest Mn neighbors have a distance of about 2.95\,\AA{}, and the exchange is apparently indirect. For direct exchange, one would expect a scaling with the magnetic moments, which is not observed here. It rather oscillates with the sp electron number. A similar result has been obtained earlier on other half and full Heusler compounds \cite{Sasioglu08}. The ratio of the nearest and second nearest neighbor coupling is significantly reduced with increasing electron concentration, and the nearest neighbor interaction dominates in Mn$_2$TiSb.

The antiferromagnetic Mn-Ti interaction is only significant for the nearest neighbors. Accordingly, the interaction between Mn and Ti, which have a distance of about 2.55\,\AA{} is essentially given by direct exchange coupling and the scaling with the Ti moment corroborates this assumption.

The intra-sublattice interaction of Mn$^{1(2)}$-Mn$^{1(2)}$ exhibits a notable oscillatory behavior. In the two cases with odd valence electron number it is positive for the nearest neighbors, negative for the second, and again positive for the third nearest neighbors. For Mn$_2$TiGe with its even electron count the first two neighbors have negative and the third neighbor has positive interaction. So in the latter case, the total Mn-Mn intra-sublattice interaction is effectively antiferromagnetic.

In order to study the dependence of $J_{ij}$ on the lattice parameter as a possible explanation for the differences discussed above, additional calculations on Mn$_2$TiGe have been performed with lattice parameters of $(5.87 \pm 0.2)$\,\AA{}. This compound was chosen because of the wide (pseudo-)gap for the spin-up states, which warrants a stable total magnetic moment and minimal band structure effects over the range of $a$ used here.

The results from these calculations are given in Fig. \ref{Fig4} (b). Obviously, the changes here are rather subtle and can not account for the the large differences discussed above. However, we note a reduction of the nearest neighbor Mn$^{1(2)}$-Mn$^{2(1)}$  interaction and of the Ti mediated second nearest Mn$^{1(2)}$-Mn$^{2(1)}$ neighbor. Meanwhile, the Mn-Ti interaction increases, in agreement with increased Mn and Ti moments.

\begin{figure}[t]
\begin{center}
\includegraphics[scale=1]{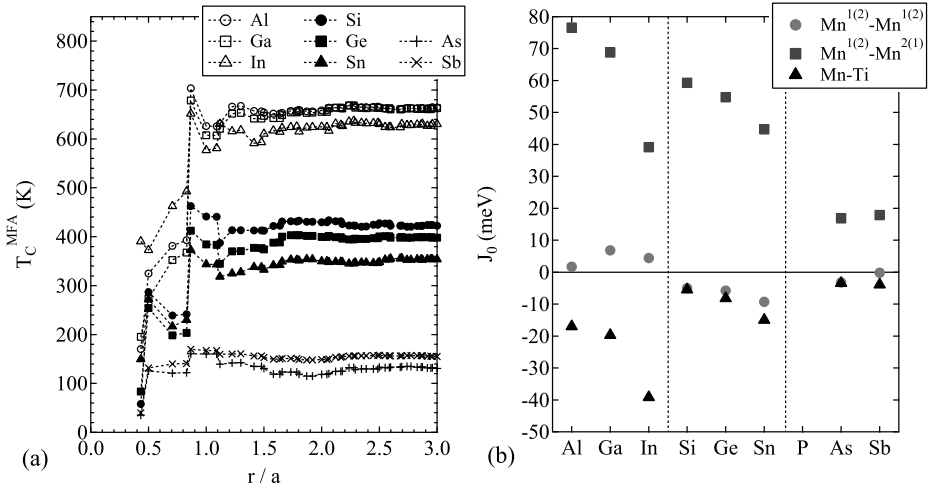}
\end{center}
\caption{\label{Fig5} (a): The Curie temperature $T_\mathrm{C}^{\mathrm{MFA}}$ in dependence on the normalized cluster radius $r/a$ taken into the summation. (b): $\bi{R}$-summed exchange coupling parameters $J_0$.}
\end{figure}

The strong confinement of the exchange interactions to a sphere with a radius of about $1.5\,a$ is reflected in the Curie temperature calculated as a function of the cluster radius which is nearly converged at $r \gtrsim 1.5\,a$, see Fig. \ref{Fig5} (a). At larger radii a weak oscillation of $T_\mathrm{C}^{\mathrm{MFA}}$ is observed, indicating long-ranged RKKY-like behaviour.

A deeper discussion of the exchange interaction is beyond the scope of this paper. However, it was recently shown for numerous half and full Heusler compounds that various exchange mechanisms---such as RKKY, superexchange and Anderson s-d mixing---contribute to the indirect exchange interactions \cite{Sasioglu08}.

The relevant contributions to the $\bi{J}_0$ matrix in Eq. (\ref{eq:multi}) are displayed in Fig. \ref{Fig5} (b). In agreement with the previous discussion it is found that the inter-sublattice interaction Mn$^{1(2)}$-Mn$^{2(1)}$ provides the largest contribution, followed by the Mn-Ti interaction, which can become as large as the Mn$^{1(2)}$-Mn$^{2(1)}$ interaction in Mn$_2$TiIn. The intra-sublattice interaction Mn$^{1(2)}$-Mn$^{1(2)}$ is generally weak, positive for Al, Ga, In, and negative for Si, Ge, Sn. All other inter- and intra-sublattice contributions are below 1\,meV. A negative intra-sublattice contribution means that the interaction acts against the ferromagnetic order on this lattice and thus reduces the Curie temperature.

To estimate the accuracy of our method for the Curie temperature determination, we calculated the Curie temperatures of some Heusler compounds at their respective experimental lattice parameters. The calculated (experimental) values are: Co$_2$MnSi 1049\,K (985\,K)\cite{Webster71}, Co$_2$TiSn 383\,K (355\,K)\cite{Majumdar05}, Mn$_2$VAl 605\,K (760\,K)\cite{Jiang01} and Mn$_2$VGa 560\,K (783\,K)\cite{Kumar08}. Further values, obtained using the same method, can be found in Ref. \cite{Thoene09}. For the Co-based ferromagnetic compounds, the calculated mean-field values are in good agreement with experiment. However, in the case of the two ferrimagnetic Mn-based compounds, the MFA Curie temperature is about 25\,\% lower than the experimental one.

\begin{table}[b]
\caption{\label{curie} Curie temperatures $T_\mathrm{C}^{\mathrm{MFA}}$ calculated in the mean-field approximation.}
\begin{indented}
\item[]\begin{tabular}{@{}l l l l l l l l l l l l l}
\br
Mn$_2$Ti\textit{Z}	&	&	Al	& Ga	& In	& 	& Si	& Ge 	& Sn	& 	& P		& As 	& Sb 	\\ \mr
$T_\mathrm{C}^{\mathrm{MFA}}$ (K)	&	& 	665	& 663	& 630	&	& 424	& 398	& 354	&	& ---	& 132	& 156	\\
\br
\end{tabular}
\end{indented}
\end{table}

Table \ref{curie} summarizes our calculated Curie temperatures. They are well above room temperature for the compounds with 21 and 22 valence electrons, but considerably lower for Mn$_2$TiAs and Mn$_2$TiSb. The Curie temperature scales roughly linear with the total magnetic moment. Within one group, the Curie temperatures are comparable, though a trend to decrease with increasing atomic number of the $Z$ component is clear for 21 and 22 valence electrons.

\begin{figure}[t]
\begin{center}
\includegraphics[scale=1]{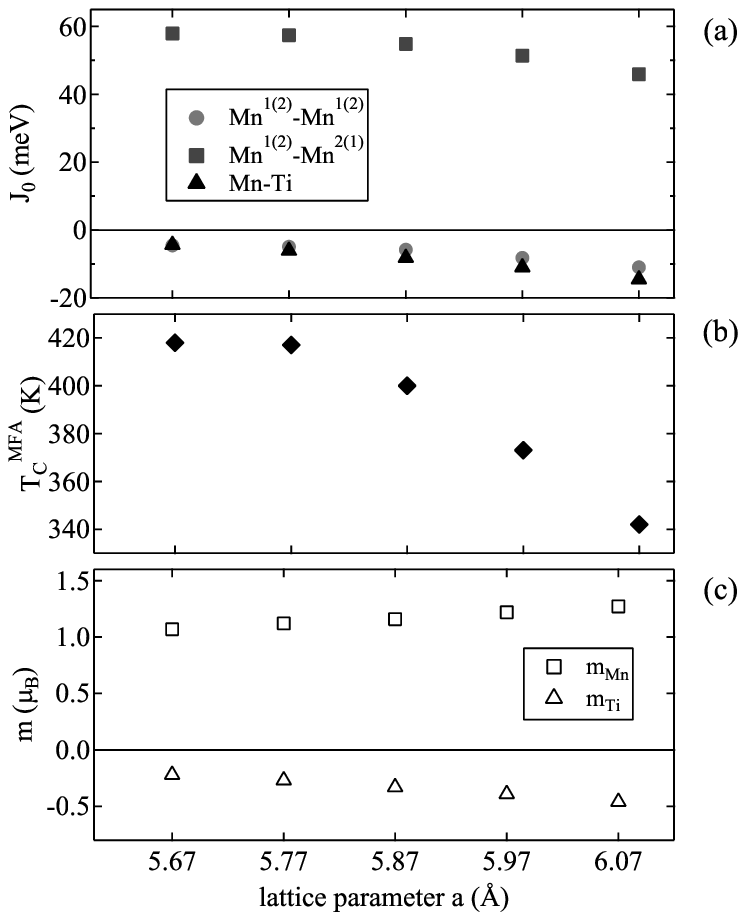}
\end{center}
\caption{\label{Fig6}Dependence of $J_0$ (a), $T_\mathrm{C}^{\mathrm{MFA}}$ (b) and magnetic moments (c) on the lattice parameter in Mn$_2$TiGe. Markers in (b) are the same as in (a). Magnetic moments in (d) are $m_\mathrm{Mn}$ (\opensquare) and $m_\mathrm{Ti}$ (\opentriangle).}
\end{figure}

The Curie temperatures of Mn$_2$TiAl, Mn$_2$TiGa and Mn$_2$TiIn are quite similar. The slightly reduced  $T_\mathrm{C}^{\mathrm{MFA}}$ of Mn$_2$TiIn is caused by the steep reduction of the Mn$^{1(2)}$-Mn$^{2(1)}$  interaction. On the other hand, a simultaneous increase of the Mn-Ti interaction stabilizes $T_\mathrm{C}^{\mathrm{MFA}}$ at a still high level. In the series Mn$_2$TiSi -- Mn$_2$TiGe -- Mn$_2$TiSn the Mn$^{1(2)}$-Mn$^{2(1)}$ decreases, but here the increase of the Mn-Ti interaction can not compensate this and hence the Curie temperature decreases. In any case, the Mn$^{1(2)}$-Mn$^{2(1)}$ interaction provides the dominant contribution to $T_\mathrm{C}^{\mathrm{MFA}}$, only in Mn$_2$TiIn  the Mn-Ti interaction is dominant. The significantly lower Curie temperature of Mn$_2$TiAs with respect to Mn$_2$TiSb can be attributed to the artificially increased lattice parameter used in the calculation.

The dependence of the exchange parameters and $T_\mathrm{C}^{\mathrm{MFA}}$ on the lattice constant was studied above for Mn$_2$TiGe. The corresponding terms of the $\bi{J}_0$ matrix, the Curie temperature and the magnetic moments are presented in Fig. \ref{Fig6} (a)-(c). A decrease of the Mn$^{1(2)}$-Mn$^{2(1)}$ interaction and simultaneously of $T_\mathrm{C}^{\mathrm{MFA}}$ with increasing $a$ is observed, although both $m_\mathrm{Mn}$ and $m_\mathrm{Ti}$ increase. Obviously, the individual moments play only a minor role in the exchange and the interatomic distances are more important. The Mn-Ti as well as the Mn$^{1(2)}$-Mn$^{1(2)}$ interactions become stronger with increasing $a$, but they nearly compensate each other. In agreement with a direct exchange coupling, the Mn-Ti interaction scales with the magnetic moments. The changes in $\bi{J}_0$ reproduce very well the changes observed in Fig. \ref{Fig5} (b) for the Si -- Ge -- Sn series.

Put in terms of a pressure dependence, we observe $\mathrm{d}T_C\,/\,\mathrm{d}p > 0$, i.e., the Curie temperature increases with increasing pressure. Kanomata \textit{et al.} proposed an empirical interaction curve for Ni$_2$Mn\textit{Z} and Pd$_2$Mn\textit{Z} full Heusler compounds that suggestes $\mathrm{d}T_C\,/\,\mathrm{d}p > 0$ for these compounds \cite{Kanomata87}. The origin of this behavior is attributed to the Mn-Mn distance and the indirect exchange between the Mn atoms, which fully carry the magnetism of the compounds. Hence, all other interactions can be neglected. A numerical confirmation by first principles of this interaction curve was given recently \cite{Sasioglu05_2}. For half-metallic Heusler compounds of type Co$_2$\textit{YZ} K\"ubler \textit{et al.} analyzed the dependence of $T_C$ on the valence electron number, which is approximately linear, and scales thus with the total magnetic moment \cite{Kuebler07}. Further it was also proposed for Co$_2$Mn\textit{Z} compounds to have $\mathrm{d}T_C\,/\,\mathrm{d}p > 0$, although the Co atom participates significantly in the exchange interactions \cite{Kurtulus05}. Experimentally this dependence on the lattice parameter was even observed for the Co$_2$Ti\textit{Z} series (with \textit{Z} = Si, Ge, Sn), where the Ti atoms have nearly vanishing magnetic moment \cite{Barth10}.

Interestingly, the magnetic moments of Mn and Ti in Mn$_2$TiGe vary within the same range as the moments for different compounds shown in Fig. \ref{Fig2}(b), while the total moment remains fixed at $2\,\mu_B$\,/\,f.u. These findings demonstrate the strong influence of the lattice parameter, while the details of the electronic structure of the \textit{Z} element are less important. Consequently, the \textit{Z} element influences the properties of the Mn$_2$Ti\textit{Z} compound mainly via its number of valence electrons and its atomic radius, which determines the equilibrium lattice parameter.

\section{Conclusion}

Our results suggest that the Mn$_2$Ti\textit{Z} Heusler compound series with \textit{Z} = Al, Ga, In, Si, Ge, Sn, P, As, Sb, can exhibit ferrimagnetism in accordance with the rule $m = N_V - 24$. Most of the compounds have large spin polarization and a spin-up gap forms above the Fermi energy. The Curie temperatures calculated within the mean-field approximation indicate that the compounds with 21 and 22 valence electrons will be ferrimagnetic at room temperature. A thorough understanding of the influence of the \textit{Z} component on the properties of the compounds has been established on the basis of \textit{ab initio} band structure and exchange coupling calculations. It was found that the pressure dependence of $T_C$ is positive, in agreement with ferromagentic full Heusler compounds. Because of their large and stable spin polarizations and their high Curie temperatures we propose in particular Mn$_2$TiSi, Mn$_2$TiGe, and Mn$_2$TiSn as candidates for spintronic applications.

\section*{Acknowledgements}
This work has been supported by the German Bundesministerium f\"ur Bildung und Forschung (BMBF) under contract number 13N9910. Helpful discussions with Prof. Andrei Postnikov are acknowledged.

\end{document}